\documentclass{PoS}

\title{Recent topics in CP violation}

\ShortTitle{CPV: Theory talk.}

\author{\speaker{Martin Jung}%
         \thanks{I would like to thank the organizers of the workshop for the invitation, as well as a very enjoyable workshop in a beautiful city.}\\
        TU Dortmund\\
        E-mail: \email{martin2.jung@tu-dortmund.de}}

\abstract{Recent topics regarding CP violation in heavy meson systems are discussed. As an introduction, the status of the Unitarity  Triangle fit and CP violation in $B$ meson mixing are briefly reviewed. 
Two topics are covered in more detail:

Penguin pollution in the ``golden mode'' $B_d\to J/\psi K$ has gained importance due to the apparent smallness of new physics effects, together with the outstanding precision expected from present and future collider experiments. A very recent analysis is presented, which yields a stronger bound for the maximal influence of penguin contributions than previous analyses and shows the corresponding uncertainty to be reducible with coming data. 

Direct CP violation in hadronic charm decays received a lot of attention lately, due to a measurement by the LHCb collaboration yielding an unexpectedly large result. While this value is certainly not generically predicted in the Standard Model, it might be possible to accommodate it nevertheless. Therefore a method is discussed to use flavour symmetries to distinguish between this possibility and new physics.
}

\FullConference{The XIth International Conference on Heavy Quarks and Leptons \\
                 11-15 June, 2012\\
                 Prague,The Czech Republic}

\begin{document}

\section{Introduction}

Roughly 40 years after its proposal \cite{Kobayashi:1973fv}, the Kobayashi-Maskawa mechanism continues to give a consistent interpretation of the available data on flavour observables and CP violation. This fact is reflected in successful fits to the Unitarity Triangle (UT) \cite{UTfits}, 
where, despite the precision data which has become available during the last decade, still no clear sign of physics beyond the Standard Model (SM) is seen. However, in the extraction of the CKM angle $\beta$ ($\phi_1$) tensions have been present (see e.g. \cite{Btaunulit}). 
The main deviation used to be between the extractions using $B\to J/\psi K$ on the one hand and $B\to \tau\nu$ on the other. However, this effect got very recently significantly reduced by the new Belle result on $B\to\tau\nu$ \cite{Adachi:2012mm}, although  the resulting world average remains above the SM expectation. 
Other puzzles, like the difference between $|V_{ub}|$ extracted from inclusive and exclusive decays or the largish $\epsilon_K$ remain, but are less significant. The important lesson from these observations is that new physics (NP) effects in the related observables have to be small. This fact, together with the bright experimental prospects, renders precision predictions for the involved observables particularly important.

Three of the observables entering the UT fit, namely the mass differences $\Delta m_{d,s}$ and the mixing induced CP asymmetry $S_{J/\psi K}$, are directly connected to $B$ meson mixing. When focusing on the possibility of NP in these systems, a dedicated analysis taking into account all available information is more appropriate. The NP influence can generally be expressed  by two complex quantities for each meson pair, parametrizing NP in the mass and rate differences. The latter is usually assumed to be SM-like, as NP entering here is strongly constrained by available data from the $B$ factories (for articles regarding this possibility see e.g. \cite{NPinGamma}). 
This assumption reduces the NP degrees of freedom to one complex parameter $\Delta_q$ for each system. Two years ago the data seemed to indicate NP in $B_s$ mixing in such a fit \cite{Lenz:2010gu}, due to an at the time apparent large mixing phase $\phi_s$, in combination with evidence for a like-sign dimuon asymmetry (LDCA) larger than the SM value \cite{Lenz:2010gu,Bmixing}. 
This situation has changed again significantly, mainly due to recent LHCb results for $\phi_s$ \cite{LHCb:2011aa,LHCb:2012ad} and the width difference $\Delta\Gamma_s$ \cite{Aaij:2012eq}. The recently updated analysis \cite{Lenz:2012az} shows a tension with the LDCA when considering NP in $\Delta m_q$ only, and has a best fit preferring SM-like $\Delta m_s$, i.e. $\Delta_s\simeq 1+0{\rm i}$. Instead, that analysis shows some indication of NP in $B_d$ mixing, driven by two effects: firstly the LDCA, the measurement of which by the D0 collaboration remains large \cite{Abazov:2011yk}. Secondly, the deviation mentioned above. The new $B\to\tau\nu$ measurement will therefore affect this analysis as well, leaving the LDCA as the only observable with a significant tension to the SM in this context\footnote{For a very recent proposal to explain this value without invoking NP in mixing, see \cite{DescotesGenon:2012kr}.}.
Furthermore, even when interpreting this result as NP, the recent BaBar result on $B\to D^{(*)}\tau\nu$ decays \cite{Lees:2012xj} would indicate rather a change in the decay amplitude than the mixing one, in contrast with one of the assumptions made in this analysis. As a result, large NP effects are ruled out here as well, underlining the importance of subleading SM effects which may mimic NP.

\section{Penguin Pollution in the Golden Modes}

The impressive precision obtained for the CKM angle $\beta$ became possible due to the fact that in the ``golden mode'', $B_d\to J/\psi K_S$, explicit calculation of the relevant matrix elements can be avoided once subleading doubly Cabibbo suppressed terms are assumed to vanish \cite{Bigi:1981qs}, in combination with a final state with a very clear experimental signature. However, given the discussion above on the size of NP effects and the precision the LHC experiments and planned next-generation $B$ factories are aiming at for this mode and related ones, a critical reconsideration of the used assumptions is mandatory. Estimates yield corrections to the famous relation $S_{J/\psi K_S}=\sin\phi_d$ of the order $\mathcal{O}(10^{-3})$, only \cite{PenEstimates}; 
it is, however, notoriously difficult to actually calculate the relevant matrix elements, and non-perturbative enhancements cannot be excluded.

To include these subleading contributions, the size of their matrix elements relative to the leading one has to be determined. An explicit calculation still does not seem feasible to an acceptable precision for the decays in question, which is why typically symmetry relations are used\footnote{For an approach using theory input to extract the $B_s$ mixing phase, see \cite{Lenz:2011zz}.}, i.e. $SU(3)$, relating up, down and strange quarks, or its subgroup $U$-spin, including only down and strange quark. These allow for accessing the unknown matrix element ratios via decays where their relative influence is larger (``control modes'') \cite{PenUspin,Faller:2008gt}. 
This method has the advantage of being a completely data-driven method, and the resulting value for the $B$ mixing phase provides improved access to NP in mixing once the SM value of this phase is determined independently. 

The main limitations of that approach were firstly the limited data for the control modes, as their rate is suppressed by the Wolfenstein parameter $\lambda$ as $\lambda^2\sim5\%$ compared to the one of $B\to J/\psi K$, and secondly corrections to the symmetry limit. The first issue was already rendered less severe by recent data from CDF and LHCb \cite{NewDataJPsiP} 
and will be resolved by LHC in combination with the planned Super Flavour Factories (SFF).  The second was addressed by a recent paper \cite{Jung:2012mp}. Here the idea is to include the symmetry-breaking corrections in a model-independent manner on a group-theoretical basis (for earlier applications of this method see e.g. \cite{Savage:1991wu,SU3breakingmodind}). 
Extending furthermore the symmetry group from $U$-spin (used in \cite{PenUspin}) 
to full $SU(3)$ then allows to relate a sufficient number of $B\to J/\psi P$ modes ($B\in\{B_u,B_d,B_s\}$ and $P\in \{\pi^+,\pi^0,K^+,K^0,\bar{K}^0\}$) to determine the parameters for the $SU(3)$ breaking as well as the penguin pollution from the fit, using mild assumptions which are mostly testable with data \cite{Jung:2012mp}.

Applying this method to presently available data for these decays \cite{NewDataJPsiP,
HFAGPDG}
shows clearly the importance of $SU(3)$-breaking effects. Even when allowing for huge values of the penguin parameters, the fit in the $SU(3)$ limit yields $\chi^2_{\rm min}/{\rm d.o.f.}=22.3(23.9)/5$, where the first number corresponds to using the former world average for the rate of $B^-\to J/\psi \pi^-$ (``dataset 1''), and the second to the new LHCb result (``dataset 2''), which yields a value about 3 standard deviations away from the former. This is why they are compared explicitly instead of averaging the results. 
Importantly, correlations to the measured branching ratios drive the shift $\Delta S=-S(B\to J/\psi K_S)+\sin\phi_d$ to relatively large values in this case, in the opposite direction of the tension observed in the UT fit.
It is furthermore interesting to note that the inclusion of neglected contributions does not improve the fit, confirming our choice to set them to zero. 
The same is true for factorizable $SU(3)$-breaking corrections, which were included in the fit for comparison purposes, only. 

In a next step, $SU(3)$-breaking contributions are included in the fit, while neglecting penguin pollution. This fit works rather well, yielding  $\chi^2_{\rm min}=9.4(6.0)$ for 7 effective degrees of freedom\footnote{\emph{Effective degrees of freedom} are defined here as number of observables minus the number of parameters which are effectively changing the fit.}. 
The best fit point yields a ratio of the larger $SU(3)$-breaking matrix element with the leading one of $19(24)\%$, which is perfectly within the expectations for this quantity. Therefore the data can be explained with the expected amount of $SU(3)$ breaking and small penguin contributions.

Performing the full fit with both additional contributions, the fit improves slightly, to $\chi^2_{\rm min}=2.8(2.3)$ for 3 effective degrees of freedom, when we refrain from applying strong restrictions on the parameter values\footnote{We do not allow for ``exchanging roles'' though, i.e. we continue to assume the leading matrix element to be the one in the $SU(3)$ limit with no penguin contributions.}. In this fit, the $SU(3)$-breaking parameters allow to accommodate the pattern of branching ratios, while the penguin contributions are mainly determined by the CP and isospin asymmetries. The central values of the penguin parameters still tend to larger values than theoretically expected. This is not surprising, given the fact that the isospin asymmetry in $B\to J/\psi K$ has a central value about ten times what is naively expected, however with large uncertainties. The corresponding branching ratios are predicted to be around one standard deviation higher (lower) for $\bar{B}^0\to J/\psi \bar{K}^0\,(B^-\to J/\psi K^-)$, making an  additional measurement of their ratio important, which correspondingly is predicted to take a significantly different central value than the one presently measured.
Restricting the fit parameters  to the expected ranges, i.e. at most $SU(3)$ breaking of $r_{SU(3)}=40\%$, and a ratio of the penguin matrix element with the leading one of $r_{\rm pen}=50\%$, shows a preference for dataset~2, where the minimal $\chi^2$ remains basically unchanged, while for dataset~1 it approximately doubles. The new result for $BR(B^-\to J/\psi\pi^-)/BR(B^-\to J/\psi K^-)$ obtained by LHCb seems therefore favoured by this fit. While  it is too early to draw conclusions, this observation demonstrates once more the importance of precise branching ratio measurements in this context.

For both datasets, the shift $\Delta S$ now tends again to positive values, thereby lowering the corresponding tension in the UT fit. It is however still compatible with zero, in agreement with the above observation of a reasonable fit without penguin terms. The obtained ranges read
\begin{eqnarray}
\Delta S_{J/\psi K}^{\rm set\,1} &=& [0.001,0.005] ([-0.004,0.011])\,,\mbox{\quad and }\\
\Delta S_{J/\psi K}^{\rm set\,2} &=& [0.004,0.011] ([-0.003,0.012])\,,
\end{eqnarray}
for $68\%$ ($95\%$)~CL, respectively, where the preferred sign change compared to the $SU(3)$ limit is due to relaxed correlations between $S(B\to J/\psi \pi^0)$ and the branching ratios in the fit, because of the additional contributions. This underlines the necessity to treat $SU(3)$ breaking model-independently.
Note that $S(B_d\to J/\psi\pi^0)$ is predicted to lie below the present central value of the measurement, thereby supporting the Belle result \cite{Lee:2007wd} over the BaBar one \cite{Aubert:2008bs}, which indicates a very large value for this observable. 
\begin{figure}
\begin{center}
\includegraphics[width=6.8cm]{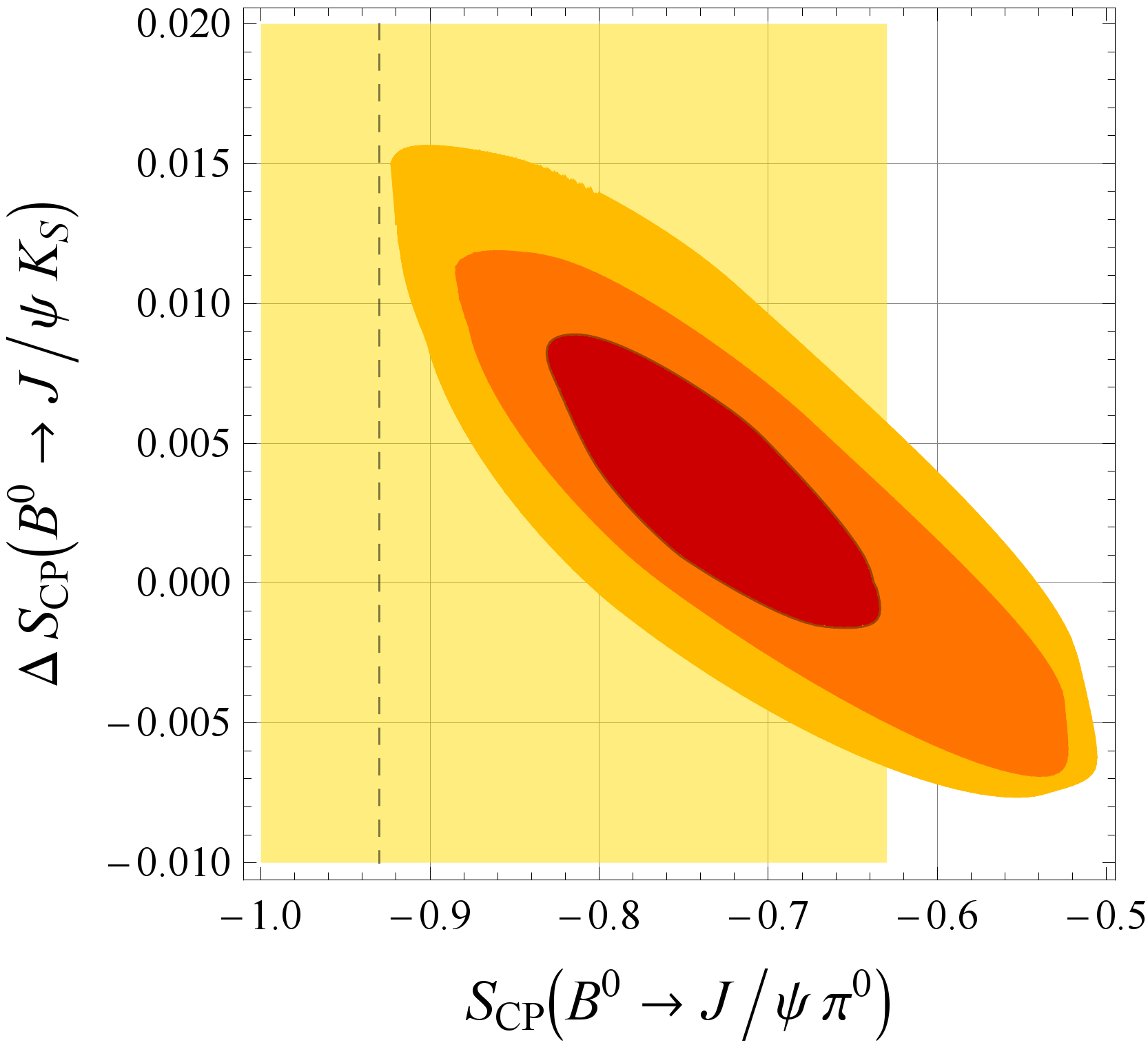}\hfill \includegraphics[width=6.8cm]{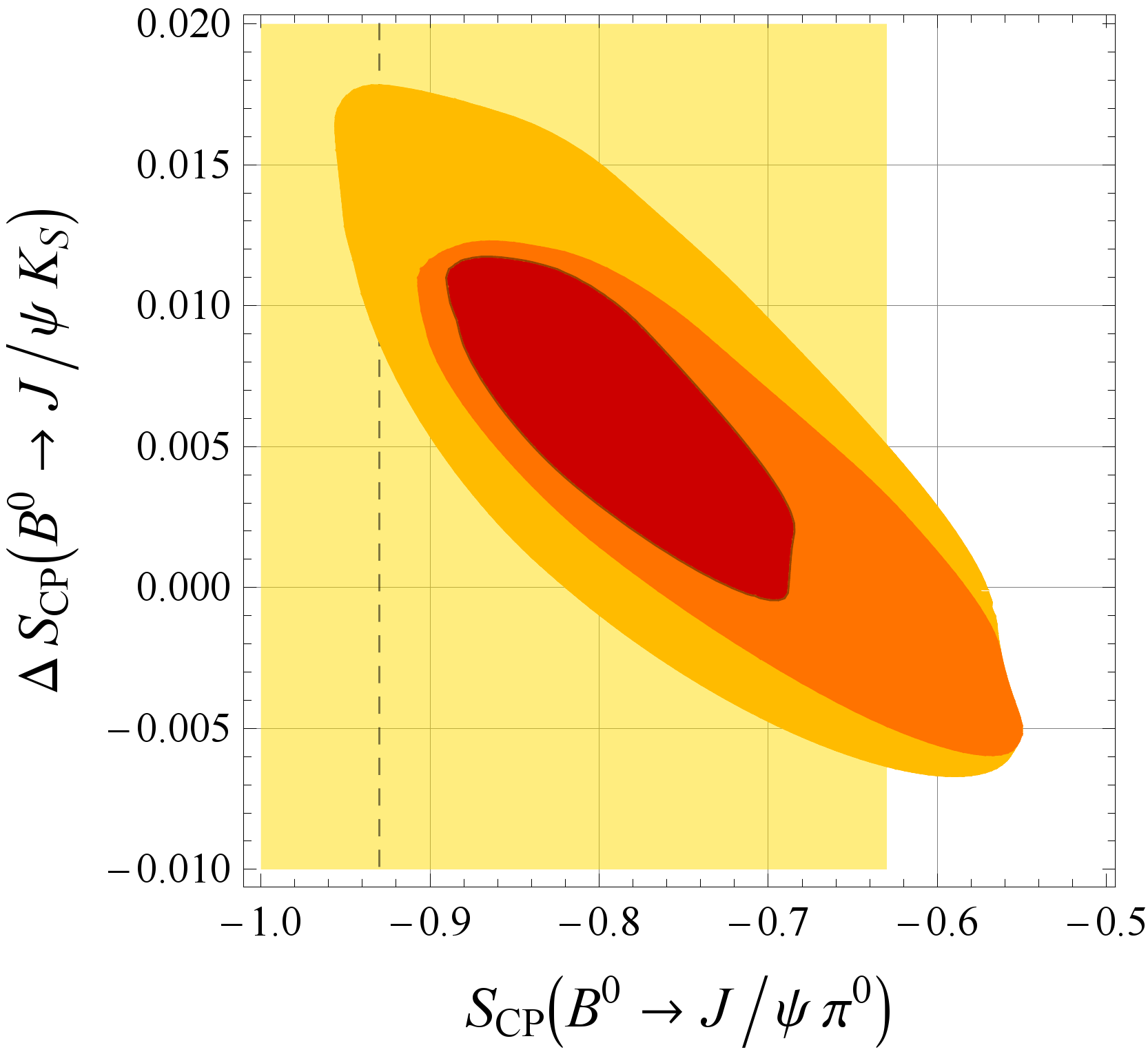}
\caption{\label{fig::resultsfullfita} Fit results for datasets 1 (left) and 2 (right), for $\Delta S$ versus $S_{\rm CP}(B^0\to J/\psi \pi^0)$, including all available data. The inner areas correspond to $68\%$ CL and $95\%$ CL with $r_{SU(3)}=40\%$ and $r_{\rm pen}=50\%$. The outer one is shown for illustration purposes, only, and corresponds to $95\%$ CL when allowing for up to $r_{SU(3)}=60\%$ and $r_{\rm pen}=75\%$. The light yellow area indicates the 2-$\sigma$ range of the $S(B^0\to J/\psi \pi^0)$ average, the dashed line its central value. Figure taken from  \cite{Jung:2012mp}.}
\end{center}
\end{figure}
These findings are illustrated in Fig.~\ref{fig::resultsfullfita}.
The same fit allows to predict the so far unmeasured CP asymmetries in $B_s\to J/\psi K$ decays: their absolute values lie for both datasets below approximately $30\%$ at $95\%$~CL. On the one hand this allows for a crosscheck of the description in the above framework, on the other hand it is clear that a measurement with a precision of $\sim10\%$ will already yield a significant additional constraint on the model parameters. Especially the dependence on the (already weak) theory assumptions will be further reduced with such a measurement \cite{Jung:2012mp}.

The mixing phase is extracted as $\phi_d^{\rm fit}=0.74\pm0.03$ (equal for both datasets), which is to be compared with $\phi_{d,{\rm naive}}^{\rm SM}=0.73\pm0.03$ when using the naive relation without penguin contributions. The inclusion of the correction therefore yields the same precision, but induces a shift of the central value. The same is true for future data, as shown in \cite{Jung:2012mp} by the consideration of several scenarios corresponding to additional data from the LHCb and SFF experiments. This implies the corresponding error to be reducible, and therefore ensures the golden mode to keep its special position among flavour observables.

In principle, the same approach can be used to constrain penguin pollution in the other ``golden mode'', $B_s\to J/\psi\phi$. Technical difficulties are the fact that the $\phi$ meson does not belong to a single representation, and the more complicated structure of the final state. The latter is also complicating the experimental analysis; so far only the $B\to J/\psi K^*$ decays have been measured, which are $b\to s$ transitions as well. If the $b\to d$ modes can be measured sufficiently precise to control the penguin pollution as well as the $SU(3)$ breaking is subject to further studies.

\section{Direct CP Violation in Hadronic Charm Decays}

The important role of the charm quark in flavour physics was established already with its discovery in in 1974 \cite{CharmDisc},
following its prediction in 1970 \cite{Glashow:1970gm}. Nowadays it remains special, due to a combination of reasons: Experimental accessibility, together with its intermediate position in the spectrum between the heavy beauty and the light strange quark (a.o. making it very well suited for 	lattice studies) and the fact that it is the only up-type quark with oscillating mesons make it a unique laboratory for understanding QCD as well as for NP searches. In particular, the corresponding information is complementary to that from the down-type quark systems.

However, the charm quark's unique position in the spectrum is also the main reason why its hadronic decays pose a severe problem for theoretical physics:
A clear hierarchy between the involved scales is absent, making most theoretical methods used for light and $B$ meson systems inapplicable, and thereby precise theoretical predictions scarce. Direct CP violation in these decays was considered to be an exception from that rule, as from its parametric suppression alone it is seen to be tiny\footnote{See however \cite{Golden:1989qx} for an early discussion of the possibility of sizable CP asymmetries in the charm system.}, $a_{\rm CP}^{\rm dir}\sim {\rm Im}(V_{cb}V_{ub}^*/(V_{cs}V_{us}^*)\sim 6\times10^{-4}$, multiplying factors which are supposed to be smaller than one. Therefore the recent LHCb measurement of $\Delta A_{\rm CP}$ \cite{Aaij:2011in}, the difference of the CP asymmetries in $D^0\to K^+K^-$ and $D^0\to \pi^+\pi^-$, confirmed by CDF \cite{Collaboration:2012qw} and very recently  also Belle \cite{Ko:2012xx}, was reason for excitement. This quantity is mainly given by the difference $\Delta a_{\rm CP}^{\rm dir}$ of the corresponding direct CP asymmetries, as the single CP asymmetries can be written in excellent approximation as the sum of the direct and the indirect one \cite{Grossman:2006jg}, the latter being universal. A small, experiment-dependent contribution from the indirect CP asymmetry remains, due to differences in the tagging for the kaons and pions. Correcting for this, the average of the three results above reads \cite{Ko:2012xx}
\begin{equation}
\Delta a_{\rm CP, exp}^{\rm dir}=(-0.68\pm0.15)\%\,,
\end{equation}
implying a combined significance of almost $5\sigma$. 

The question if this value is compatible with the SM has been answered differently in the literature (see e.g. \cite{Isidori:2011qw,BrodFranco,
Feldmann:2012js,Brod:2012ud,Mannel:2012hb}), and many NP models have been examined with respect to their capability to explain the observed value naturally, e.g. in \cite{Isidori:2011qw,Feldmann:2012js,Hochberg:2011ru,NPinD:2012}.
The problem can be reduced to the question if large enhancement factors can occur in a penguin matrix element relative to a tree one for this system. This ratio would naively be expected to be much smaller than one, hence the expectation for a tiny CP asymmetry in the SM. For the $B$ meson system this ratio is indeed small, while in the kaon system it shows an ``anomaly'', typically referred to as the $\Delta I=1/2$ rule. This enhancement, while not fully understood theoretically, is not expected to appear in a similar form for $D$ mesons, given the absence of strong hierarchies necessary for such an effect. However, as an actual calculation does not appear feasible, some enhancement cannot be excluded.

Given this unsatisfactory situation, the inclusion of additional data is necessary. To this aim, there have been different recent proposals, discussing e.g. rare or multibody $D$ decays, but also information from other systems like top decays or electric dipole moments, see e.g. \cite{Mannel:2012hb,Hochberg:2011ru,AltObs}. 
Alternatively, other non-leptonic decays can be related to those in question by flavour symmetries \cite{CharmSU3Original}, typically either $SU(3)$ or $U$-spin.
However, the symmetry limit is known to yield a bad fit to the data in this case, making the inclusion of corrections necessary. Especially the observed ratio $BR(D^0\to K^+K^-)/BR(D^0\to\pi^+\pi^-)\sim 2.8$, which is expected to be approximately one in the symmetry limit, questions the applicability of these methods generally. Such a large ratio can be explained in part by factorizable $SU(3)$ breaking in terms of form factors and decay constants, but not fully. However, in \cite{Savage:1991wu} it has been pointed out that already a general breaking of $\sim20-30\%$ (which is expected for these symmetries) can suffice to explain this ratio, simply because this correction applies on the amplitude level, and is of opposite sign for the decays in question. This has been confirmed in recent publications by explicit fits to the data, which included symmetry-breaking corrections to the $U$-spin limit \cite{Feldmann:2012js,Brod:2012ud,Hiller:2012xm}. While in this framework the data can be  accommodated, it does not allow for predictions for other observables. For that, the symmetry group has to be extended to $SU(3)$ (see e.g. \cite{SU3recent} for recent publications). The breaking in this case can be chosen such that it conserves isospin, for which no strong breaking is visible \cite{Savage:1991wu,Hinchliffe:1995hz}. However, due to the more complicated group structure of the breaking terms, the corresponding fit cannot be carried out in full generality. 
A reasonable fit requires the inclusion of several $SU(3)$-breaking matrix elements, at least one of which needs to be from a higher representation \cite{Hiller:2012xm}, at variance with one possibility discussed in the literature assuming the dominance of small representations (see e.g. \cite{Kwong:1993ri,Pirtskhalava:2011va}).
To what extent this approach provides the possibility  to distinguish between the SM and different NP scenarios, will be discussed in detail in \cite{Hiller:2012xm}.

\section{Conclusions}
CP violation studies in heavy meson systems remain a very active field, and one of the main paths to discover NP. The general picture remains consistent with the KM mechanism as the only source of low-energy CP violation; in fact, the fits have improved very recently due to a new measurement for $B\to\tau\nu$. However, some tensions remain which require clarification in the future. The fits for NP in the $B_{d,s}$ systems show again mainly consistency with the SM, apart from the open question of the LDCA for which it is quite difficult to find a consistent theoretical explanation. 

This -- in many ways unexpected -- situation requires a more precise knowledge of the corresponding SM expectations, as potential small NP contributions will compete with subleading SM ones. The ``golden modes'' $B_d\to J/\psi K$ and $B_s\to J/\psi\phi$ are examples where subleading contributions can affect the extraction of the mixing phase. For $B_d\to J/\psi K$, a new approach to control them has been advocated, allowing to take into account $SU(3)$ corrections model-independently, which were shown to affect the procedure severely. The main result is a new limit, $|\Delta S_{J/\psi K}|\lesssim 0.01$ ($95\%$~CL), which can additionally be improved by coming data.

Last winter's measurement of a non-vanishing direct CP asymmetry in the $D$ system is a major achievement, and triggered many theoretical analyses. Nevertheless, a clear interpretation of this result in terms of the SM or NP is spoiled by the lack of a theoretical method to calculate the relevant hadronic matrix elements. The main strategy to decide that question is relating this observation to other observables. One option are flavour symmetries, which connect the decays in question to other hadronic $D$ decays. An important observation is that the corresponding data for the decay rates can be accommodated with a reasonable amount of $SU(3)$ breaking. This confirms the applicability of the method, thereby providing another path to clarify the issue of SM versus NP in hadronic charm decays.

In conclusion, the apparent smallness of NP effects in flavour observables poses challenges to both theory and experiment. On the experimental side they are met by several high-luminosity collider experiments, both running and under construction, allowing for unprecedented precision. Also on the theory side the challenges are answered, by new strategies and adapting known ones to higher precision. Together, these developments make for an exciting way ahead.

\section*{Acknowledgements}
This work is supported by the Bundesministerium f\"ur Bildung und Forschung (BMBF).

\section*{Note added}
After this article was finished, the LHCb collaboration announced the first measurement of $B_s\to J/\psi K^*$ \cite{Aaij:2012nh}. This measurement will allow for further insight into $B\to J/\psi V$ decays, $V$ denoting a vector meson, allowing for the first time for an application of the approach presented in \cite{Faller:2008gt} to data.


\begin{thebibliography}{99}
\bibitem{Kobayashi:1973fv}
  M.~Kobayashi and T.~Maskawa,
  Prog.\ Theor.\ Phys.\  {\bf 49} (1973) 652.

\bibitem{UTfits}
  J.~Charles {\it et al.}  [CKMfitter Group Collaboration],
  Eur.\ Phys.\ J.\ C {\bf 41} (2005) 1
  [hep-ph/0406184]. Updated results and plots available at: {\tt http://ckmfitter.in2p3.fr} .
%
  M.~Ciuchini, G.~D'Agostini, E.~Franco, V.~Lubicz, G.~Martinelli, F.~Parodi, P.~Roudeau and A.~Stocchi,
  JHEP {\bf 0107} (2001) 013
  [hep-ph/0012308]. Updated results and plots available at: {\tt http://www.utfit.org} .

\bibitem{Btaunulit}
  O.~Deschamps,
  arXiv:0810.3139 [hep-ph].
%
  T.~Feldmann, M.~Jung and T.~Mannel,
  JHEP {\bf 0808} (2008) 066
  [arXiv:0803.3729 [hep-ph]].
%
  M.~Bona {\it et al.}  [UTfit Collaboration],
  Phys.\ Lett.\ B {\bf 687} (2010) 61
  [arXiv:0908.3470 [hep-ph]].
%
  E.~Lunghi and A.~Soni,
  Phys.\ Rev.\ Lett.\  {\bf 104} (2010) 251802
  [arXiv:0912.0002 [hep-ph]].
%
  J.~Charles, O.~Deschamps, S.~Descotes-Genon, R.~Itoh, H.~Lacker, A.~Menzel, S.~Monteil and V.~Niess {\it et al.},
  Phys.\ Rev.\ D {\bf 84} (2011) 033005
  [arXiv:1106.4041 [hep-ph]].

\bibitem{Adachi:2012mm}
  I.~Adachi {\it et al.}  [Belle Collaboration],
  arXiv:1208.4678 [hep-ex].

 
\bibitem{NPinGamma}
  A.~Dighe, A.~Kundu and S.~Nandi,
  Phys.\ Rev.\ D {\bf 82} (2010) 031502
  [arXiv:1005.4051 [hep-ph]].
%
  C.~W.~Bauer and N.~D.~Dunn,
  Phys.\ Lett.\ B {\bf 696} (2011) 362
  [arXiv:1006.1629 [hep-ph]].
%
  S.~Oh and J.~Tandean,
  Phys.\ Lett.\ B {\bf 697} (2011) 41
  [arXiv:1008.2153 [hep-ph]].
 %
  C.~Bobeth and U.~Haisch,
  arXiv:1109.1826 [hep-ph].

\bibitem{Lenz:2010gu}
  A.~Lenz, U.~Nierste, J.~Charles, S.~Descotes-Genon, A.~Jantsch, C.~Kaufhold, H.~Lacker and S.~Monteil {\it et al.},
  Phys.\ Rev.\ D {\bf 83} (2011) 036004
  [arXiv:1008.1593 [hep-ph]].

\bibitem{Bmixing}  
  A.~Lenz and U.~Nierste,
  JHEP {\bf 0706} (2007) 072
  [hep-ph/0612167].
%
  V.~M.~Abazov {\it et al.}  [D0 Collaboration],
  Phys.\ Rev.\ D {\bf 82} (2010) 032001
  [arXiv:1005.2757 [hep-ex]].
%
  M.~Ciuchini, E.~Franco, V.~Lubicz, F.~Mescia and C.~Tarantino,
  JHEP {\bf 0308} (2003) 031
  [hep-ph/0308029].

\bibitem{LHCb:2011aa}
  R.~Aaij {\it et al.}  [LHCb Collaboration],
  Phys.\ Rev.\ Lett.\  {\bf 108} (2012) 101803
  [arXiv:1112.3183 [hep-ex]].

\bibitem{LHCb:2012ad}
  R.~Aaij {\it et al.}  [LHCb Collaboration],
  Phys.\ Lett.\ B {\bf 713} (2012) 378
  [arXiv:1204.5675 [hep-ex]].
  
\bibitem{Aaij:2012eq}
  R.~Aaij {\it et al.}  [LHCb Collaboration],
  Phys.\ Rev.\ Lett.\  {\bf 108} (2012) 241801
  [arXiv:1202.4717 [hep-ex]].

\bibitem{Lenz:2012az}
  A.~Lenz, U.~Nierste, J.~Charles, S.~Descotes-Genon, H.~Lacker, S.~Monteil, V.~Niess and S.~T'Jampens,
  Phys.\ Rev.\ D {\bf 86} (2012) 033008
  [arXiv:1203.0238 [hep-ph]].

\bibitem{Abazov:2011yk}
  V.~M.~Abazov {\it et al.}  [D0 Collaboration],
  Phys.\ Rev.\ D {\bf 84} (2011) 052007
  [arXiv:1106.6308 [hep-ex]].

\bibitem{DescotesGenon:2012kr}
  S.~Descotes-Genon and J.~F.~Kamenik,
  arXiv:1207.4483 [hep-ph].

\bibitem{Lees:2012xj}
  J.~P.~Lees {\it et al.}  [BaBar Collaboration],
  Phys.\ Rev.\ Lett.\  {\bf 109} (2012) 101802
  [arXiv:1205.5442 [hep-ex]].


\bibitem{Bigi:1981qs}
  I.~I.~Y.~Bigi and A.~I.~Sanda,
  Nucl.\ Phys.\ B {\bf 193} (1981) 85.

\bibitem{PenEstimates}  
  H.~Boos, T.~Mannel and J.~Reuter,
  Phys.\ Rev.\ D {\bf 70} (2004) 036006
  [hep-ph/0403085].
  %
  H.~-n.~Li and S.~Mishima,
  JHEP {\bf 0703} (2007) 009
  [hep-ph/0610120].
  %
  M.~Gronau and J.~L.~Rosner,
  Phys.\ Lett.\ B {\bf 672} (2009) 349
  [arXiv:0812.4796 [hep-ph]].

\bibitem{Lenz:2011zz}
  A.~J.~Lenz,
  Phys.\ Rev.\ D {\bf 84} (2011) 031501
  [arXiv:1106.3200 [hep-ph]].

\bibitem{PenUspin}
  R.~Fleischer,
  Eur.\ Phys.\ J.\ C {\bf 10} (1999) 299
  [hep-ph/9903455].
 %
  M.~Ciuchini, M.~Pierini and L.~Silvestrini,
  Phys.\ Rev.\ Lett.\  {\bf 95} (2005) 221804
  [hep-ph/0507290].
 %
  M.~Ciuchini, M.~Pierini and L.~Silvestrini,
  arXiv:1102.0392 [hep-ph].
%
  S.~Faller, M.~Jung, R.~Fleischer and T.~Mannel,
  Phys.\ Rev.\ D {\bf 79} (2009) 014030
  [arXiv:0809.0842 [hep-ph]].
%
\bibitem{Faller:2008gt}
  S.~Faller, R.~Fleischer and T.~Mannel,
  Phys.\ Rev.\ D {\bf 79} (2009) 014005
  [arXiv:0810.4248 [hep-ph]].

\bibitem{NewDataJPsiP}
  T.~Aaltonen {\it et al.}  [CDF Collaboration],
  Phys.\ Rev.\ D {\bf 83} (2011) 052012
  [arXiv:1102.1961 [hep-ex]].
%
  RAaij {\it et al.}  [LHCb Collaboration],
  Phys.\ Lett.\ B {\bf 713} (2012) 172
  [arXiv:1205.0934 [hep-ex]].
%
  R.~Aaij {\it et al.}  [LHCb Collaboration],
  Phys.\ Rev.\ D {\bf 85} (2012) 091105
  [arXiv:1203.3592 [hep-ex]].
  
\bibitem{Jung:2012mp}
  M.~Jung,
  Phys.\ Rev.\ D {\bf 86} (2012) 053008
  [arXiv:1206.2050 [hep-ph]].

\bibitem{Savage:1991wu}
  M.~J.~Savage,
  Phys.\ Lett.\ B {\bf 257} (1991) 414.

\bibitem{SU3breakingmodind}
  M.~Gronau, O.~F.~Hernandez, D.~London and J.~L.~Rosner,
  Phys.\ Rev.\ D {\bf 52} (1995) 6356
  [hep-ph/9504326].
 %
  M.~Jung and T.~Mannel,
  Phys.\ Rev.\ D {\bf 80} (2009) 116002
  [arXiv:0907.0117 [hep-ph]].

\bibitem{HFAGPDG}
  Y.~Amhis {\it et al.}  [Heavy Flavor Averaging Group Collaboration],
  arXiv:1207.1158 [hep-ex], and online update at http://www.slac.stanford.edu/xorg/hfag.
%
  J.~Beringer {\it et al.}  (Particle Data Group),
  Phys.\ Rev.\ D {\bf 86} (2012) 010001.
  
\bibitem{Lee:2007wd}
  S.~E.~Lee {\it et al.}  [Belle Collaboration],
  Phys.\ Rev.\ D {\bf 77} (2008) 071101
  [arXiv:0708.0304 [hep-ex]].

\bibitem{Aubert:2008bs}
  B.~Aubert {\it et al.}  [BABAR Collaboration],
  Phys.\ Rev.\ Lett.\  {\bf 101} (2008) 021801
  [arXiv:0804.0896 [hep-ex]].

\bibitem{CharmDisc}
  J.~E.~Augustin {\it et al.}  [SLAC-SP-017 Collaboration],
  Phys.\ Rev.\ Lett.\  {\bf 33} (1974) 1406.
%
  J.~J.~Aubert {\it et al.}  [E598 Collaboration],
  Phys.\ Rev.\ Lett.\  {\bf 33} (1974) 1404.

\bibitem{Glashow:1970gm}
  S.~L.~Glashow, J.~Iliopoulos and L.~Maiani,
  Phys.\ Rev.\ D {\bf 2} (1970) 1285.

\bibitem{Golden:1989qx}
  M.~Golden and B.~Grinstein,
  Phys.\ Lett.\ B {\bf 222} (1989) 501.
  
\bibitem{Aaij:2011in}
  R.~Aaij {\it et al.}  [LHCb Collaboration],
  Phys.\ Rev.\ Lett.\  {\bf 108} (2012) 111602
  [arXiv:1112.0938 [hep-ex]].
  
\bibitem{Collaboration:2012qw}
  T.~Aaltonen {\it et al.}  [CDF Collaboration],
  Phys.\ Rev.\ Lett.\  {\bf 109} (2012) 111801
  [arXiv:1207.2158 [hep-ex]].
  
\bibitem{Ko:2012xx}
  B.R.~Ko for the Belle Collaboration,
  ``Direct CP violation in charm at Belle'',
  Talk given at ICHEP 2012, Melbourne.
  Average by Marco Gersabeck.

\bibitem{Grossman:2006jg}
  Y.~Grossman, A.~L.~Kagan and Y.~Nir,
  Phys.\ Rev.\ D {\bf 75} (2007) 036008
  [hep-ph/0609178].

\bibitem{Isidori:2011qw}
  G.~Isidori, J.~F.~Kamenik, Z.~Ligeti and G.~Perez,
  Phys.\ Lett.\ B {\bf 711} (2012) 46
  [arXiv:1111.4987 [hep-ph]].

\bibitem{BrodFranco}
  J.~Brod, A.~L.~Kagan and J.~Zupan,
  Phys.\ Rev.\ D {\bf 86} (2012) 014023
  [arXiv:1111.5000 [hep-ph]].
%
  E.~Franco, S.~Mishima and L.~Silvestrini,
  JHEP {\bf 1205} (2012) 140
  [arXiv:1203.3131 [hep-ph]].

\bibitem{Feldmann:2012js}
  T.~Feldmann, S.~Nandi and A.~Soni,
  JHEP {\bf 1206} (2012) 007
  [arXiv:1202.3795 [hep-ph]].

\bibitem{Brod:2012ud}
  J.~Brod, Y.~Grossman, A.~L.~Kagan and J.~Zupan,
  JHEP {\bf 1210} (2012) 161
  [arXiv:1203.6659 [hep-ph]].
  
\bibitem{Mannel:2012hb}
  T.~Mannel and N.~Uraltsev,
  arXiv:1205.0233 [hep-ph].



\bibitem{Hochberg:2011ru}
  Y.~Hochberg and Y.~Nir,
  Phys.\ Rev.\ Lett.\  {\bf 108} (2012) 261601
  [arXiv:1112.5268 [hep-ph]].
  
\bibitem{NPinD:2012}
  A.~N.~Rozanov and M.~I.~Vysotsky,
  arXiv:1111.6949 [hep-ph].
%
  G.~F.~Giudice, G.~Isidori and P.~Paradisi,
  JHEP {\bf 1204} (2012) 060
  [arXiv:1201.6204 [hep-ph]].
%
  W.~Altmannshofer, R.~Primulando, C.~-T.~Yu and F.~Yu,
  JHEP {\bf 1204} (2012) 049
  [arXiv:1202.2866 [hep-ph]].
%
  C.~-H.~Chen, C.~-Q.~Geng and W.~Wang,
  Phys.\ Rev.\ D {\bf 85} (2012) 077702
  [arXiv:1202.3300 [hep-ph]].
%
  G.~Hiller, Y.~Hochberg and Y.~Nir,
  Phys.\ Rev.\ D {\bf 85} (2012) 116008
  [arXiv:1204.1046 [hep-ph]].
%
  B.~Keren-Zur, P.~Lodone, M.~Nardecchia, D.~Pappadopulo, R.~Rattazzi and L.~Vecchi,
  Nucl.\ Phys.\ B {\bf 867} (2013) 429
  [arXiv:1205.5803 [hep-ph]].
%
  C.~-H.~Chen, Chao-Qiang and W.~Wang,
  arXiv:1206.5158 [hep-ph].
%
  C.~Delaunay, J.~F.~Kamenik, G.~Perez and L.~Randall,
  arXiv:1207.0474 [hep-ph].
%
  Y.~Grossman, A.~L.~Kagan and J.~Zupan,
  Phys.\ Rev.\ D {\bf 85} (2012) 114036
  [arXiv:1204.3557 [hep-ph]].

\bibitem{AltObs}
  I.~I.~Bigi and A.~Paul,
  JHEP {\bf 1203} (2012) 021
  [arXiv:1110.2862 [hep-ph]].
%
  G.~Isidori and J.~F.~Kamenik,
  Phys.\ Rev.\ Lett.\  {\bf 109} (2012) 171801
  [arXiv:1205.3164 [hep-ph]].
%
  I.~I.~Bigi,
  arXiv:1206.4554 [hep-ph].

\bibitem{CharmSU3Original}
  M.~B.~Einhorn and C.~Quigg,
  Phys.\ Rev.\ D {\bf 12} (1975) 2015.
  R.~L.~Kingsley, S.~B.~Treiman, F.~Wilczek and A.~Zee,
  Phys.\ Rev.\ D {\bf 11} (1975) 1919.
%
  G.~Altarelli, N.~Cabibbo and L.~Maiani,
  Phys.\ Lett.\ B {\bf 57} (1975) 277.
%
  J.~F.~Donoghue and L.~Wolfenstein,
  Phys.\ Rev.\ D {\bf 15} (1977) 3341.
%
  C.~Quigg,
  Z.\ Phys.\ C {\bf 4} (1980) 55.
%
  L.~-L.~Chau,
  Phys.\ Rept.\  {\bf 95} (1983) 1.
%

\bibitem{Hiller:2012xm}
  G.~Hiller, M.~Jung and S.~Schacht,
  arXiv:1211.3734 [hep-ph].

\bibitem{Hinchliffe:1995hz}
  I.~Hinchliffe and T.~A.~Kaeding,
  Phys.\ Rev.\ D {\bf 54} (1996) 914
  [hep-ph/9502275].

\bibitem{Kwong:1993ri}
  W.~Kwong and S.~P.~Rosen,
  Phys.\ Lett.\ B {\bf 298} (1993) 413.
  
\bibitem{Pirtskhalava:2011va}
  D.~Pirtskhalava and P.~Uttayarat,
  Phys.\ Lett.\ B {\bf 712} (2012) 81
  [arXiv:1112.5451 [hep-ph]].

\bibitem{SU3recent}
  B.~Bhattacharya, M.~Gronau and J.~L.~Rosner,
  Phys.\ Rev.\ D {\bf 85} (2012) 054014
  [arXiv:1201.2351 [hep-ph]].
  H.~-Y.~Cheng and C.~-W.~Chiang,
  Phys.\ Rev.\ D {\bf 85} (2012) 034036
v   [Erratum-ibid.\ D {\bf 85} (2012) 079903]
  [arXiv:1201.0785 [hep-ph]].
%
  H.~-Y.~Cheng and C.~-W.~Chiang,
  Phys.\ Rev.\ D {\bf 86} (2012) 014014
  [arXiv:1205.0580 [hep-ph]].
  
\bibitem{Aaij:2012nh}
  RAaij {\it et al.}  [LHCb Collaboration],
  Phys.\ Rev.\ D {\bf 86} (2012) 071102
  [arXiv:1208.0738 [hep-ex]].
\end{thebibliography}
\end{document}